\input{aipcheck}

\documentclass[
    ,final            
  ]
  {aipproc}

\layoutstyle{8x11double}

\begin{document}

\title{Nuclear PDF for neutrino and charged lepton data}


\author{K.~Kova\v{r}\'{\i}k}{
  address={Institute for Theoretical Physics, Karlsruhe Institute of Technology, Karlsruhe, D-76128,Germany}
}

\begin{abstract}
Neutrino Deep Inelastic Scattering (DIS) on nuclei is an essential process to constrain the strange quark parton 
distribution functions (PDF) in the proton. The critical component on the way to using the neutrino DIS data in a proton PDF 
analysis is understanding the nuclear effects in parton distribution functions. We parametrize these effects by nuclear 
parton distribution functions (NPDF). Here we compare results from two analysis
of NPDF both done at next-to-leading order in QCD. The first uses neutral current charged-lepton 
$(\ell^{\pm}A)$ Deeply Inelastic Scattering (DIS) and Drell-Yan data for several nuclear
targets and the second uses neutrino-nucleon DIS data. We compare the nuclear corrections factors
($F_{2}^{Fe}/F_{2}^{D}$) for the charged-lepton data with other
results from the literature. In particular, we compare and contrast
fits based upon the charged-lepton DIS data with those using neutrino-nucleon DIS data.
\end{abstract}

\maketitle

%
\section{Introduction}
Parton distribution functions (PDFs) are an indispensable part of any prediction involving hadrons in the initial state.
This is the reason why many groups perform and regularly update global analysis of PDFs for protons \cite{Ball:2009mk, 
Martin:2009iq, Nadolsky:2008zw,JimenezDelgado:2008hf} and nuclei \cite{Hirai:2007sx, Eskola:2009uj,deFlorian:2003qf}. 
Although not often emphasized, nuclear effects are present also in the proton PDFs analysis as a number of experimental data is 
taken on nuclear targets. Mostly though, the nuclear targets used in the proton analysis, are made of light nuclei where nuclear effects are generally 
small. An important exception is the neutrino DIS data which is taken on heavy nuclei such as iron or lead and is sensitive to 
the strange quark content of the proton. A knowledge of the strange quark PDF has an influence on precise measurements at the LHC such as 
$W$- or $Z$-boson production. 

In order to make use of the neutrino DIS data to constrain the strange quark PDF, we 
have to apply a nuclear correction factor which can be obtained either from a specific model of nuclear 
interactions \cite{Kulagin:2004ie} or from an analysis of nuclear parton distribution functions (NPDF) based on experimental data. 

Here, we present a framework for a global analysis of nuclear PDFs at next-to-leading order in QCD closely related to the CTEQ 
framework for proton PDFs. We analyze and compare the nuclear correction factor obtained from the usual charged lepton DIS and 
Drell-Yan (DY) data to the one from the neutrino DIS data.   
\section{Nuclear PDF}
The global NPDF framework, we use to analyze charged lepton DIS and DY data and neutrino DIS data, was introduced in 
\cite{Schienbein:2009kk}. The parameterizations of the parton distributions in bound protons at the input scale of $Q_0=1.3 {\rm GeV}$	
\begin{equation}
x\, f_{k}(x,Q_{0}) = c_{0}x^{c_{1}}(1-x)^{c_{2}}e^{c_{3}x}(1+e^{c_{4}}x)^{c_{5}}\,,\label{eq:input1}
\end{equation}
where $k=u_{v},d_{v},g,\bar{u}+\bar{d},s,\bar{s}$ and
\begin{equation}	
\bar{d}(x,Q_{0})/\bar{u}(x,Q_{0}) = c_{0}x^{c_{1}}(1-x)^{c_{2}}+(1+c_{3}x)(1-x)^{c_{4}}\,,\label{eq:input2} 
\end{equation}
are a generalization of the parton parameterizations in free protons used in the CTEQ proton analysis \cite{Pumplin:2002vw}. To account for a variety 
of nuclear targets, the coefficients $c_k$ are generalized to functions of the nucleon number $A$
\begin{equation}
c_{k}\to c_{k}(A)\equiv c_{k,0}+c_{k,1}\left(1-A^{-c_{k,2}}\right),\ k=\{1,\ldots,5\}\,.\label{eq:Adep}
\end{equation}
The proton PDF in this framework are obtained as a limit $A\rightarrow 1$ and are held fixed at values obtained in the analysis 
\cite{Pumplin:2002vw}. From the input distributions, we can construct the PDFs for a general $(A,Z)$-nucleus 
\begin{equation}
f_{i}^{(A,Z)}(x,Q)=\frac{Z}{A}\ f_{i}^{p/A}(x,Q)+\frac{(A-Z)}{A}\ f_{i}^{n/A}(x,Q),\label{eq:pdf}
\end{equation}
where we relate the distributions of a bound neutron, $f_{i}^{n/A}(x,Q)$, to those of a proton by isospin symmetry. 

We performed a global analysis of nuclear charged lepton DIS and DY data within this framework, determining the $A$-dependence of the 
parameters $c_k(A)$. In the analysis, we applied the same standard kinematic cuts $Q>2 {\rm GeV}$ and $W>3.5 {\rm GeV}$ as in 
\cite{Pumplin:2002vw} and obtain a fit with $\chi^{2}/{\rm dof}$ of 0.946 to 708 data points with 32 free parameters (for 
further details see \cite{Schienbein:2009kk}). 

\begin{figure}[h!]
\includegraphics[width=.45\textwidth]{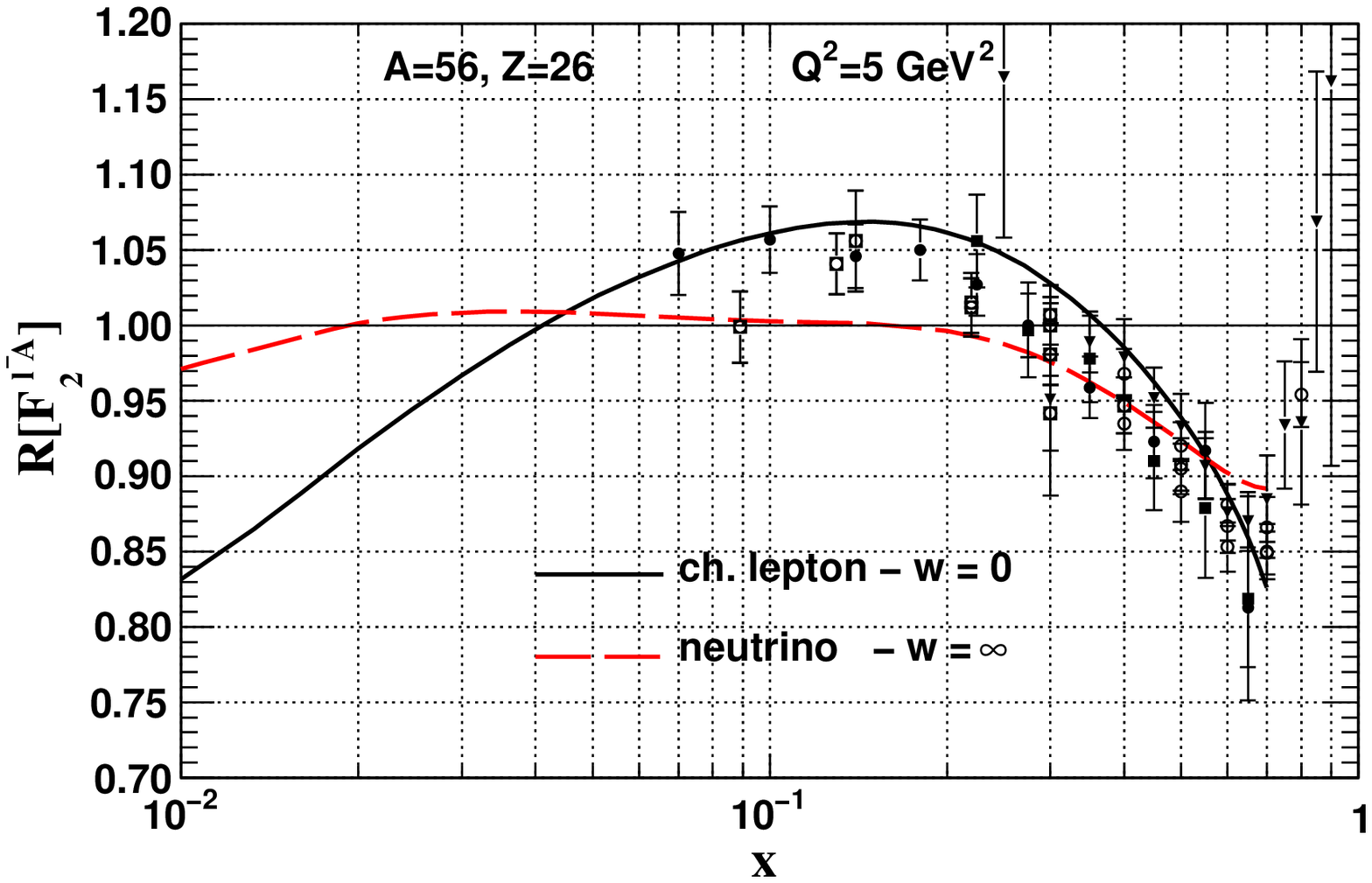}
\includegraphics[width=.45\textwidth]{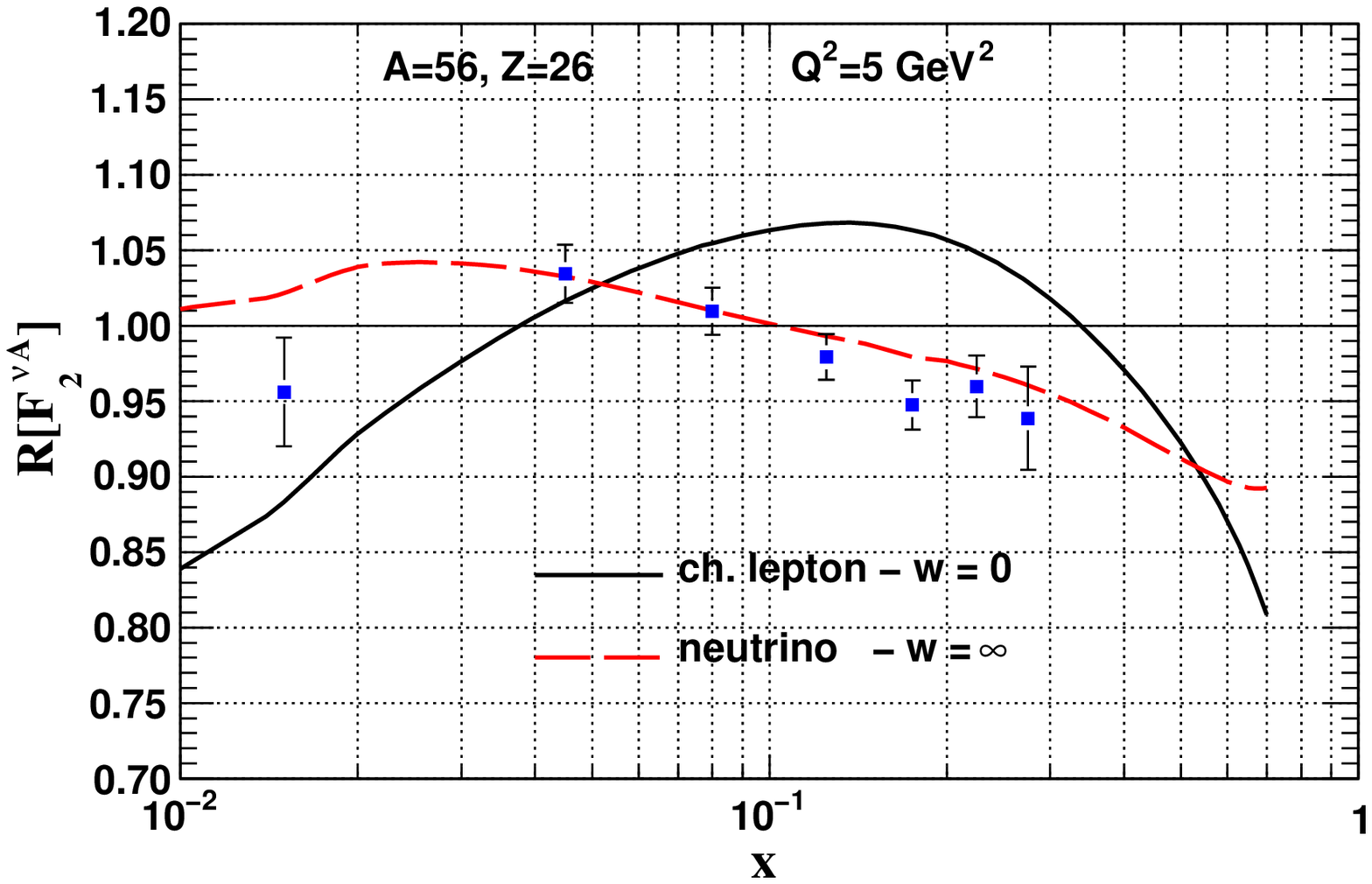}
  \caption{Nuclear correction factors $R_{NC}^{e,\mu}(F_2;x,Q^2)$ (left) and $R_{CC}^{e,\mu}(F_2;x,Q^2)$ (right) for global fits to charged lepton DIS and DY data (solid line) and to neutrino DIS cross-section data (dashed line) at the scale $Q^2=5 {\rm GeV}^2$.}\label{fig:f2-0}
\end{figure}

The nuclear effects extracted in the form of NPDF are usually presented in the form of nuclear correction 
factors. We focus on two nuclear correction factors related either to the DIS structure 
function $F_2$ in the charged-current (CC) $\nu A$ process
\begin{equation}
R_{CC}^\nu(F_2;x,Q^2)\simeq \frac{d^A+\bar{u}^A+\ldots}{d^{A,0}+\bar{u}^{A,0}+\ldots}\,,
\label{eq:rcc}
\end{equation}
or to the DIS structure function $F_2$ in the neutral-current (NC) $l^\pm A$ process
\begin{eqnarray}\nonumber
&& R_{NC}^{e,\mu}(F_2;x,Q^2)\simeq \\ && 
\frac{[d^A + \bar{d}^A + \ldots]+ 4 [u^A + \bar{u}^A+\ldots]}{[d^{A,0} + \bar{d}^{A,0} + \ldots]
+4 [u^{A,0} + \bar{u}^{A,0}+\ldots]}\,.
\label{eq:rnc}
\end{eqnarray}
The superscript $`0'$ stands for using the free nucleon PDFs $f_i^p(x,Q)$ and $f_i^n(x,Q)$
in Eq.\ (\ref{eq:pdf}) instead of the bound nucleon distributions $f_i^{p/A}(x,Q)$ and $f_i^{n/A}(x,Q)$. 

In Fig.~\ref{fig:f2-0} (solid line), we show how the result of our global analysis of charged lepton data translates into these 
nuclear correction factors and how it compares to experimental data. As first observed in \cite{Schienbein:2007fs}, the 
$R_{CC}^\nu(F_2;x,Q^2)$ correction factor calculated using Eq.~\ref{eq:rcc} with parton densities from the fit to the charged 
lepton nuclear data, does not describe the NuTeV data well which raises the question if including neutrino 
DIS data in the global analysis corrects this behavior without spoiling the $R_{NC}^{e,\mu}(F_2;x,Q^2)$ correction factor which 
fits the charged lepton DIS and DY data well.

\section{NPDF from neutrino DIS data}
To investigate the apparent discrepancy between the predicted nuclear correction factor $R_{CC}^\nu(F_2;x,Q^2)$ from the fit to charged lepton data and the neutrino charged current DIS data, we have 
set up a global analysis where we used exclusively the neutrino DIS cross-section data coming from NuTeV and Chorus experiments 
taken on iron and lead respectively. Here we applied the same kinematic cuts as in the first analysis of the charged lepton data 
and we obtain a fit to 3134 neutrino DIS cross-section data points with $\chi^{2}/{\rm dof}$ of 1.33 with 34 free parameters (for 
further details see \cite{Kovarik:2010uv}).
\begin{table}
\begin{tabular}{lccc}
\hline 
$w$  & $ \chi^{2}_{l^{\pm}A}$ (/pt)  & $\chi^{2}_{\nu A}$ (/pt)  & total $\chi^{2}$(/pt)\tabularnewline
\hline 
$0$    & 638 (0.90)   & -  & 638 (0.90) \tabularnewline
$1/7$   & 645 (0.91)    & 4710 (1.50)  & 5355 (1.39) \tabularnewline
$1/2$    & 680 (0.96)   & 4405 (1.40)  & 5085 (1.32) \tabularnewline
$1$   & 736 (1.04)      & 4277 (1.36)  & 5014 (1.30) \tabularnewline
$\infty$  & -   & 4192 (1.33)  & 4192 (1.33) \tabularnewline
\hline
\end{tabular}
\caption{Summary table of a family of compromise fits. \label{tab:compr} }
\end{table}

As was expected, the global fit to neutrino DIS data describes the data for the charged current nuclear correction factor 
$R_{CC}^\nu(F_2;x,Q^2)$ well and does a poor job to describe the neutral current correction factor especially at low and 
intermediate Bjorken $x$. We see that using one or the other data sets produces different nuclear correction factors 
$R_{NC}^{e,\mu}(F_2;x,Q^2)$ and $R_{CC}^\nu(F_2;x,Q^2)$. A question arises if there are such nuclear correction factors which 
would be in agreement with both charged lepton and neutrino data, for example using a combined set of charged lepton and neutrino DIS data.
Analyzing both data sets in a combined global analysis runs into the problem of imbalance of number of data points 
between the two data sets. This would automatically mean that the neutrino data would be favored just based on the amount of 
data. Therefore, we introduce an artificial parameter, the weight of the neutrino data set $w$, 
\begin{equation}
\chi^{2}=\sum_{l^{\pm}A\ {\rm data}}\chi_{i}^{2}\ +\!\!\sum_{\nu A\ {\rm data}}w\,\chi_{i}^{2}\ ,\label{eq:chi2}
\end{equation}
\begin{figure}[t]
\includegraphics[width=.45\textwidth]{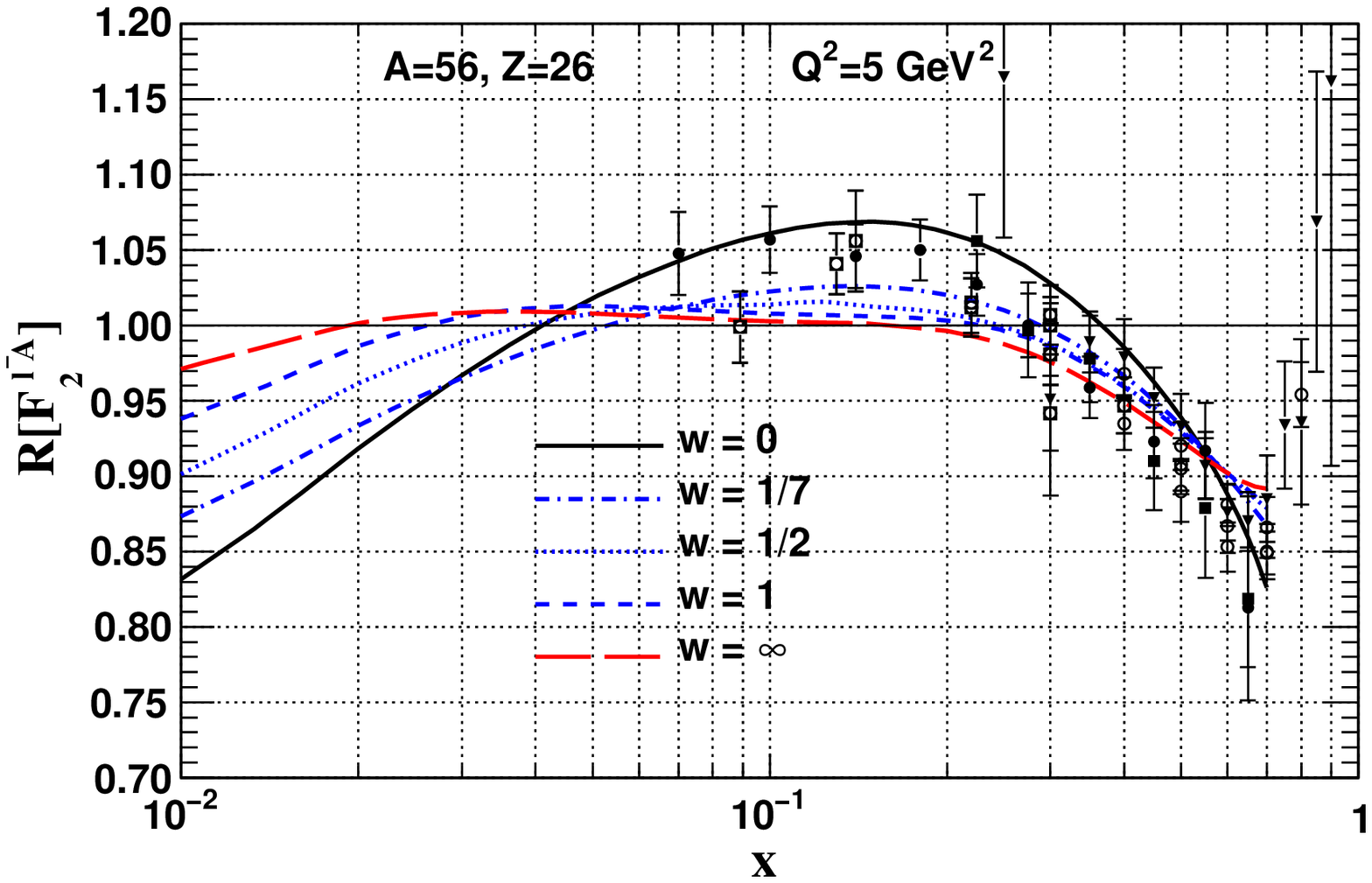}
\includegraphics[width=.45\textwidth]{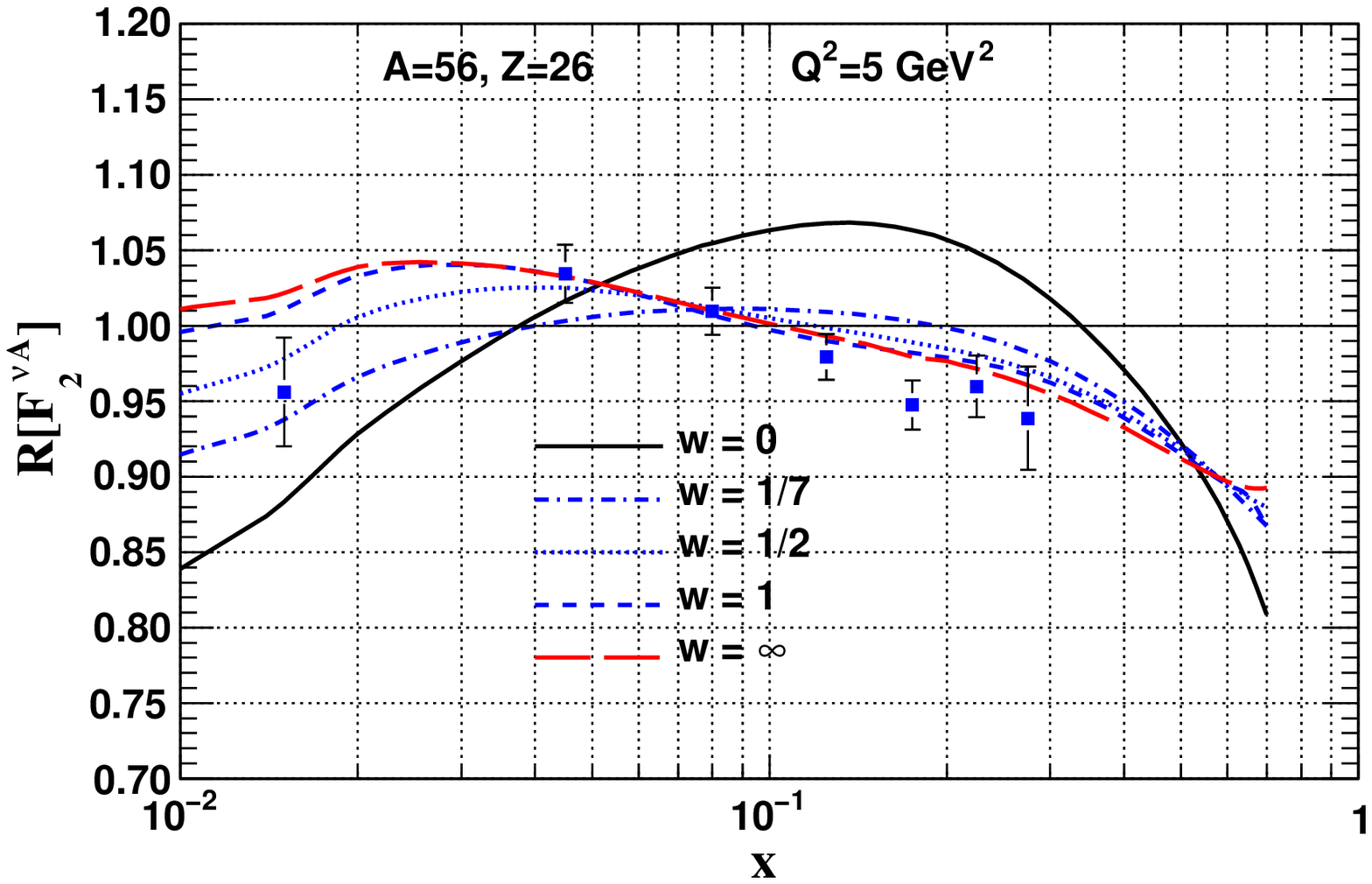}
  \caption{Nuclear correction factors $R_{NC}^{e,\mu}(F_2;x,Q^2)$ (left) and $R_{CC}^{e,\mu}(F_2;x,Q^2)$ (right) for compromise fits with different weights of the neutrino DIS data at the scale $Q^2=5 {\rm GeV}^2$.}\label{fig:f2-1}
\end{figure}
to interpolate between the two different global fits ($w=0$ results in the fit to charged lepton data only and $w=\infty$ stands 
symbolically for the fit only to neutrino data). Varying the weight $w$, we try to find a compromise fit which would describe both charged lepton and neutrino data 
well. We list the resulting $\chi^2$ for the compromise fits with weights $w=0,1/7,1/2,1,\infty$ in Tab.~\ref{tab:compr} and we show the nuclear correction factors 
$R_{NC}^{e,\mu}(F_2;x,Q^2)$ and $R_{CC}^\nu(F_2;x,Q^2)$ for the same family of compromise fits in Fig.~\ref{fig:f2-1}.

Indeed we see in Fig.~\ref{fig:f2-1} that the fits with $w=1/7,1/2,1$ interpolate well between the two extreme cases $w=0$ and $w=\infty$. In order to decide on how well the compromise fits describe the data we use the $\chi^2$ goodness-of-fit criterion introduced and used in \cite{Stump:2001gu,Martin:2009iq}. 
We consider a fit a good compromise if its $\chi^2$ for both data subsets, the charged lepton DIS and DY data and the neutrino DIS data, is within 90\% confidence level of the fits to only charged lepton or neutrino data. 

We define the 90\% percentile $\xi_{90}$ used to define the 90\% confidence level, by
\begin{equation}\label{xi90}
	\int_0^{\xi_{90}}P(\chi^2,N)d\chi^2 = 0.90\,,
\end{equation}
where $N$ is the number of degrees of freedom and $P(\chi^2, N)$ is the probability distribution 
\begin{equation}\label{chi2dist}
	P(\chi^2,N) = \frac{(\chi^2)^{N/2-1}e^{-\chi^2/2}}{2^{N/2}\Gamma(N/2)}\,.
\end{equation}
We can assign a 90\% confidence level error band to the $\chi^2$ of the fits to the charged lepton DIS and DY data and to the neutrino DIS data 
\begin{equation}\label{lnuA90}
	\chi^2_{l^\pm A} = 638+ 45.6,\qquad \chi^2_{\nu A} = 4192+ 138.
\end{equation}
Comparing the results of the compromise fits with different weights, listed in Tab.~\ref{tab:compr}, we see that none of the compromise fits 
are compatible with both 90\% confidence level limits given in Eq.\ref{lnuA90}. As detailed in \cite{Kovarik:2010uv}, not even relaxing 
the condition to compare against the 99\% confidence level limit helps to finding a suitable compromise fit. Moreover, we show in \cite{Kovarik:2010uv}
that the effect is related to the precise neutrino DIS data from NuTeV.
\section{Conclusion}
After performing a thorough global NPDF analysis of the combined charged lepton and neutrino data, we find that there is no good compromise description of 
both the data sets simultaneously. The differences are most pronounced in the low and intermediate $x$ regions where the neutrino DIS (NuTeV) 
do not show a strong shadowing effect as the charged lepton data do. The inability to describe all data by one consistent framework indicates the existence of non-universal nuclear effects or unexpectedly large higher-twist effects.  
here

\bibliographystyle{aipproc}   

\bibliography{npdf}

\end{document}